\documentclass[twocolumn,showpacs,preprintnumbers,amsmath,amssymb,APSl,prd,nofootinbib,superscriptaddress]{revtex4-1}

\usepackage{dcolumn}
\usepackage{bm}
\usepackage{ifpdf}
\usepackage{hyperref}
\usepackage{bm}
\usepackage{xcolor,color,graphicx,graphics}
\usepackage[spanish,english]{babel}
\usepackage[latin1]{inputenc}
\usepackage[OT1]{fontenc}
\usepackage{latexsym,amssymb,amsmath,amsfonts}
\usepackage{makeidx}
\usepackage{epsfig,subfigure}
\usepackage{natbib}
\usepackage{epstopdf}
\usepackage{mathrsfs}
\usepackage{hyperref}
\hypersetup{colorlinks=true, linkcolor=blue, citecolor=green}
\usepackage{enumerate}

\definecolor{red}{rgb}{1,0,0}

\def\+{^\dagger}

\def\<{\leftarrow}
\def\>{\rightarrow}

\def\({\left(}
\def\){\right)}

 \def\n{\nu}

\newcommand{\bi}{\begin{itemize}} 				\newcommand{\ei}{\end{itemize}}
\newcommand{\benu}{\begin{enumerate}} 		\newcommand{\enu}{\end{enumerate}}
\newcommand{\bd}{\begin{dinglist}{0}}     \newcommand{\ed}{\end{dinglist}}
\newcommand{\bfig}{\begin{figure}[htbp]}  \newcommand{\efig}{\end{figure}}
        			
\newcommand{\bc}{\begin{center}} 				  \newcommand{\ec}{\end{center}}
\newcommand{\be}{\begin{equation}} 				\newcommand{\ee}{\end{equation}}
\newcommand{\bsub}{\begin{subequations}}  \newcommand{\esub}{\end{subequations}}
\newcommand{\ben}{\begin{eqnarray}} 			\newcommand{\een}{\end{eqnarray}}
\newcommand{\ba}[1]{\begin{array}{#1}} 		\newcommand{\ea}{\end{array}}
\newcommand{\bea}{\begin{equation}\begin{array}{rcl}}
\newcommand{\eea}{\end{array}\end{equation}}

\begin{document}
\title{Charged BTZ-type solutions in Eddington-inspired Born-Infeld gravity}

\author{Merce Guerrero} \email{merguerr@ucm.es}
\affiliation{Departamento de F\'isica Te\'orica and IPARCOS,
	Universidad Complutense de Madrid, E-28040 Madrid, Spain}	
	
\author{Gerardo Mora-P\'{e}rez} \email{moge@alumni.uv.es}
\affiliation{Departamento de F\'{i}sica Te\'{o}rica and IFIC, Centro Mixto Universidad de Valencia - CSIC.
	Universidad de Valencia, Burjassot-46100, Valencia, Spain}

\author{Gonzalo J. Olmo} \email{gonzalo.olmo@uv.es}
\affiliation{Departamento de F\'{i}sica Te\'{o}rica and IFIC, Centro Mixto Universidad de Valencia - CSIC.
	Universidad de Valencia, Burjassot-46100, Valencia, Spain}
\affiliation{Departamento de F\'isica, Universidade Federal da
	Para\'\i ba, 58051-900 Jo\~ao Pessoa, Para\'\i ba, Brazil}
	
	\author{Emanuele Orazi} \email{orazi.emanuele@gmail.com}
\affiliation{ International Institute of Physics, Federal University of Rio Grande do Norte,
Campus Universit\'ario-Lagoa Nova, Natal-RN 59078-970, Brazil}
\affiliation{Escola de Ciencia e Tecnologia, Universidade Federal do Rio Grande do Norte, Caixa Postal 1524, Natal-RN 59078-970, Brazil}

\author{Diego Rubiera-Garcia} \email{drubiera@ucm.es}
\affiliation{Departamento de F\'isica Te\'orica and IPARCOS,
	Universidad Complutense de Madrid, E-28040 Madrid, Spain}

\date{\today}
\begin{abstract}
We construct an axially symmetric solution of Eddington-inspired Born-Infeld gravity coupled to an electromagnetic field in $2+1$ dimensions including a (negative) cosmological constant term. This is achieved by using a recently developed mapping procedure that allows to generate solutions in certain families of metric-affine gravity theories starting from a known seed solution of General Relativity, which in the present case corresponds to the electrically charged Banados-Teitelboim-Zanelli (BTZ) solution. We discuss the main features of the new configurations, including the modifications to the ergospheres and horizons, the emergence of wormhole structures, and the consequences for the regularity (or not) of these space-times via geodesic completeness.
\end{abstract}

\maketitle

\section{Introduction}

The Banados-Teitelboim-Zanelli (BTZ) metric describes an axially symmetric solution of the 2+1 dimensional Einstein-Maxwell field equations of General Relativity (GR) with a negative cosmological constant term. Its original interest lied on the fact that it allows for the existence of a negative-mass Anti-de Sitter space disconnected from the spectrum of black holes by a mass gap \cite{Banados:1992wn,Banados:1992gq}. Such a state has not event horizon covering it but curvature scalars are everywhere finite, rendering it as a kind of regular naked object. This finding sparked the investigation of smoothing out black hole singularities by replacing the innermost region by de Sitter cores \cite{Dymnikova,Ansoldi:2008jw,Lemos:2011dq,Maeda:2021jdc}, the role of the BTZ solution within the AdS/CFT correspondence \cite{Chamblin:1999tk}, or the analysis of its quasi-normal modes \cite{Cardoso:2001hn}. Moreover, the BTZ solution has been extended to include an electric charge \cite{Carlip:1995qv,Cataldo:1996ue} and further generalized to non-linear matter fields \cite{Cataldo:2000ns,Hendi:2012zz,Gonzalez:2021vwp} and to the quantum realm \cite{Emparan:2020znc}.

On the other hand, it is known since a long time ago that an ultraviolet completion of GR requires the introduction of higher-order curvature terms suppressed by inverse powers of Planck's mass \cite{UVC1,UVC2}. An exceedingly large number of proposal has been considered on how exactly to implement this extension \cite{Review1,Review2,Review3,Review4}, and which do not need (and often indeed not) to agree on what the new theory should do regarding the modification of the classical GR solutions. A widespread expectation of such generalized solutions, however, is that they will be able to resolve the problem with space-time singularities inside black holes (see e.g. \cite{Senovilla:2014gza} for an in-depth discussion of the nature of these singularities and the theorems predicting them). Therefore, the finding of exact three dimensional solutions of theoretical/physical interest of these new theories is a source of great interest in the literature \cite{Oliva:2009ip,Alkac:2016xlr,Hendi:2017mgb,Gurses:2019wpb,Tang:2019jkn,Konoplya:2020ibi,Bueno:2021krl}. Since such extensions typically introduce higher-order field equations and/or strongly increase their non-linear character,  several procedures have been developed to shortcut the structure of such field equations. One such methods is the Newman-Janis one \cite{Newman:1965tw,Newman:1965my}, upon which many new solutions including rotating and non-singular black holes have been found, and their observational signatures  discussed \cite{Bambi:2013ufa,Toshmatov:2014nya,Tsukamoto:2017fxq,Shaikh:2019fpu,Hennigar:2020drx,Mazza:2021rgq,Shaikh:2021yux}.

Another generating method has been recently developed for theories of gravity formulated in metric-affine spaces, where metric and affine connection are regarded as independent entities \cite{Bahamonde:2021akc}. The current research has identified a family of such theories, dubbed as Ricci-based gravities (RBGs), that, while being capable to successfully pass solar system tests and the gravitational wave observations so far, allow for the introduction of an Einstein frame representation of their field equations \cite{Afonso:2018bpv}. Using this frame it is possible to establish a correspondence (or mapping) between GR coupled to a set of matter fields described by some Lagrangian density, and a given RBG coupled to the same kind of fields but described by a different Lagrangian density \cite{Afonso:2018mxn,Afonso:2018hyj}. This way, once a seed solution on the GR side is known, one can find the counterpart on the RBG side via purely algebraic transformations. In vacuum, this mapping trivializes and one recovers GR and its solutions, but in the presence of matter sources new exact solutions of these theories have been found in axially symmetric scenarios \cite{Guerrero:2020azx,Shao:2020weq} and even in setups without symmetries \cite{Olmo:2020fnk}, allowing to explore the phenomenology of metric-affine theories in a much more efficient way \cite{Ali:2021psk}.

The main aim of this work is to use the mapping method above in order to generate the counterpart of the BTZ solution in a member of the RBG class, the so-called Eddington-inspired Born-Infeld (EiBI) gravity, which has attracted a great deal of attention in the literature (for a review of the inception of this theory and its many applications see \cite{BeltranJimenez:2017doy}). We shall work out the specific shape of the mapping for electromagnetic fields in $2+1$ dimensions by establishing the correspondence between the matter actions on each frame, finding that GR coupled to a Maxwell field maps into EiBI gravity coupled to a Born-Infeld-type of non-linear electrodynamics. Using this fact, we shall use a seed solution of the former, as given by the (charged) BTZ solution, to generate a new solution of the latter via the mapping. We will double-check the validity of this solution  by re-deriving it using an anisotropic fluid analogy for the electromagnetic field. Then we shall study the main features of the newly found configurations, which are qualitatively different depending on the sign of the EiBI parameter. In particular, we will discuss the ergospheres and horizons of these solutions, the structure of the radial function, and its impact on the innermost region regarding the issue of geodesic completeness.

\section{Mapping procedure of RBG theories}

To start with our derivations, let us consider the RBG family of theories in $D=n+1$ space-time dimensions as defined by the following action
\begin{equation} \label{eq:RBGaction}
S_{RBG}=\int d^D x \sqrt{-g} \mathcal{L}_G(g_{\mu\nu},R_{(\mu\nu)})  + \mathcal{S}_m(g_{\mu\nu},\psi_m) \ ,
\end{equation}
where $g$ is the determinant of the space-time metric $g_{\mu\nu}$ and the gravitational Lagrangian density $\mathcal{L}_G$ must be constructed out of traces of the object ${M^\mu}_{\alpha} \equiv g^{\mu\nu}R_{(\nu\alpha)}$ (to yield a scalar object), where $R_{(\mu\nu)}(\Gamma)$ is the symmetric part of the Ricci tensor\footnote{This symmetrization requirement is introduced to safeguard the theory against ghost-like instabilities without introducing additional torsional pieces in order to eat up such ghosts \cite{BeltranJimenez:2019acz,Jimenez:2020dpn}. To lighten the notation, from now on we drop the parenthesis.} of the affine connection, $\Gamma \equiv \Gamma_{\mu\nu}^{\lambda}$, the latter being independent of the metric (Palatini or metric-affine formalism). As for the matter sector $\mathcal{S}_m=\int d^D x \sqrt{-g}\mathcal{L}_m(g_{\mu\nu},\psi_m)$, it depends only on the metric and the matter fields $\psi_m$, but not on the connection, which is free of ambiguities as long as we consider minimally coupled bosonic fields \cite{Afonso:2017bxr}.
It has been shown elsewhere (see e.g. \cite{Afonso:2018bpv}) that in 3+1 space-time dimensions, the field equations of the RBG family can be consistently reduced to the Einstein equations when the metric is suitably redefined and new interactions in the matter sector are included. In order to extend this statement to the 2+1 dimensional case, we have to revisit all the steps that led to the RBG-GR correspondence paying attention to the dependence of the geometrical quantities on the number of space-time dimensions.

\subsection{Metric compatibility of RBG theories }

As far as the field equations coming from the variation of the RBG action \eqref{eq:RBGaction} are concerned, no dependence on the dimension is manifest. Indeed, one can explicitly check that the variation with respect to the connection leads to the following kinematic constraint:
\be
\frac{1}{\sqrt{|g|}}\nabla_\mu\left(\sqrt{|g|} {\cal Z}^{\beta\lambda}\right) \delta^{\mu\nu}_{\alpha\lambda} = S^\nu{}_{\alpha\lambda}{\cal Z}^{\beta\lambda} + 2 S^\lambda{}_{\lambda\mu}{\cal Z}^{\beta\lambda}\delta^{\mu\nu}_{\alpha\lambda}\label{eq:ConnFieldEq}
\ee
where the tensor
\be
{\cal Z}^{\mu\nu} \equiv \frac{\partial{{\cal L}_G}}{\partial R_{\mu\nu}}\ , \label{eq:Z-Def}
\ee
has been introduced and $S_{\mu\nu}^{\lambda}=\tfrac{1}{2}(\Gamma_{\mu\nu}^{\lambda}-\Gamma_{\nu\mu}^{\lambda})$ is the torsion tensor. Taking special care of the different factors that depend on the specific choice of a 2+1 dimensional space-time, one can follow the same steps outlined in \cite{Orazi:2020mhb}, reducing the field equations \eqref{eq:ConnFieldEq} to the following relation
\be
\nabla_\lambda {q}^{\mu\nu}  =2\left(S^\nu{}_{\lambda\alpha}q^{\mu\alpha} - S^\alpha{}_{\alpha\beta}q^{\mu\gamma}\delta^\beta_{(\lambda}\delta^\nu_{\gamma)}\right)\ ,\label{eq:ConnFieldEqMin}
\ee
where the auxiliary metric
\be
q^{\mu\nu} \equiv  \xi\left|\frac{\cal Z}{g}\right| {\cal Z}^{\mu\nu}\,,\label{eq:qDef}
\ee
has been defined in terms of the determinant $\cal Z \equiv \det{(\cal Z_{\mu\nu})}$ of the inverse of the tensor introduced in \eqref{eq:Z-Def} and an arbitrary constant $\xi$. Performing the following projective transformation
\be
\Gamma^\lambda_{\mu\nu} = \tilde\Gamma^\lambda_{\mu\nu} - S^\alpha{}_{\alpha\mu} \delta^\lambda_\nu\,,\label{eq:SpecProjTransf}
\ee
the field equations \eqref{eq:ConnFieldEqMin} assume the following manageable aspect
\be
\tilde\nabla_\lambda {q}^{\mu\nu}  = 2 \tilde S^\nu{}_{\lambda\alpha}q^{\mu\alpha}\ ,\label{eq:FinConnEq}
\ee
where from now on tildes over covariant derivatives imply that they are defined with respect to the tilded connection as defined in \eqref{eq:SpecProjTransf}. From \eqref{eq:FinConnEq}, it is straightforward to write the tilded connection in terms of the auxiliary metric to finally solve the connection field equation as follows
\be
\tilde\Gamma^\lambda{}_{(\alpha\nu)} = \frac12 q^{\lambda\mu}\left(\partial_\alpha q_{\mu\nu} + \partial_\nu q_{\alpha\mu} - \partial_\mu q_{\nu\alpha}\right) \hspace{0.1cm}; \hspace{0.1cm}  \tilde S^\lambda{}_{\alpha\nu}=0\ ,\label{MetrComp}
\ee
putting forward the possibility to describe an RBG as a (pseudo-) Riemannian geometry in terms of the auxiliary metric $q_{\mu\nu}$.

\subsection{RBG as GR theory}

In analogy with the 3+1 dimensional  case, the auxiliary metric $q_{\mu\nu}$ can be regarded as a solution of GR provided that the field equations associated to the RBG share the same matter dependence as the Einstein field equations when expressed in terms of $q_{\mu\nu}$. Therefore, we define the Einstein (or GR) frame as the theory that is based on the action
\be
S_{GR} = \frac{1}{2\kappa^2}\int d^3 x \sqrt{-q} \tilde R(\tilde\Gamma) \ , \label{eq:GR-action}
\ee
which is the metric-affine formulation of GR, where the fields and variables that are inherent to GR are characterized by a tilde. Looking at the metric field equations in the RBG and GR frames, which are given by\footnote{Note that the dependence of the gravitational Lagrangian on $R^\mu{}_\nu$ allows to use the following identity $\frac{\delta {\cal L}_G\left(R^\mu{}_\nu\right)}{\delta g^{\rho\sigma}} = \frac{\delta {\cal L}_G\left(R^\mu{}_\nu\right)}{\delta R_{\alpha\beta}}g_{\alpha\rho}R_{\beta\sigma}$.}
\be
\frac{\delta{\cal L}_G}{\delta R_{\mu\rho}} R_{\rho\nu}(\Gamma) = \frac12 T^\mu{}_\nu + \frac12 {\cal L}_G\delta^\mu_\nu\ ,\label{eq:PreEinstein}
\ee
and
\be
q^{\mu\rho} R_{\rho\nu}(\tilde\Gamma )=\kappa^2\left(\tilde T^\mu{}_\nu-\tilde T \delta^\mu_\nu\right)\ , \label{eq:AnalogueGR}
\ee
respectively, the effective description of the two theories is equivalent if the following condition holds
\be
\sqrt{-q}\left(\tilde T^\mu{}_\nu- \tilde T \delta^\mu_\nu\right) =\sqrt{-g}\left( T^\mu{}_\nu + {\cal L}_G\delta^\mu_\nu\right)\,,\label{eq:PreCond}
\ee
where the definition of the auxiliary metric \eqref{eq:qDef} has been rephrased as follows
\be
\sqrt{-q} q^{\mu\nu} = 2\kappa^2\sqrt{-g} \frac{\delta{\cal L}_G}{\delta R_{\mu\nu}}\ ,\label{eq:AuxMetrEq}
\ee
thus fixing the arbitrary constant to $\xi =\left(2\kappa^2\right)^{-2}$, where $\kappa^2=8\pi G$ is Newton's constant. From the definition of the energy-momentum tensor, $T_{\mu\nu}(g)=\tfrac{-2}{\sqrt{-g}} \tfrac{\delta \mathcal{S}_m}{\delta g^{\mu\nu}}$, and its analogue in the Einstein frame, $\tilde{T}_{\mu\nu}(q)=\tfrac{-2}{\sqrt{-q}} \tfrac{\delta \tilde{\mathcal{S}}_m}{\delta q^{\mu\nu}}$,
the condition \eqref{eq:PreCond} can be shown to provide the following recipe to relate the matter sectors on each frame:
\begin{eqnarray}
&&\sqrt{-g}\mathcal{L}_m(g,\psi) \label{eq:RBGParamtr} \\
&=&2\sqrt{-q}\left( q^{\mu\nu}\frac{\delta\tilde{\cal L}_m(q,\psi)}{\delta q^{\mu\nu}} -\tilde{\cal L}_m(q,\psi)\right)-\sqrt{-q}\mathcal{L}_G\ , \nonumber
\end{eqnarray}
if the space of solutions  is restricted by the conditions
\be
\sqrt{-g}g^{\mu\rho}\frac{\delta{\cal L}_m(g,\psi)}{\delta g^{\rho\nu}} = \sqrt{-q}{q}^{\mu\rho}\frac{\delta\tilde{\cal L}_m(q,\psi)}{\delta {q}^{\rho\nu}}\ .\label{eq:Restr}
\ee
In order for these parameterizations to be useful, the explicit expression of the gravity Lagrangian in terms of the matter fields is needed. This relation is encoded in the field equations associated to the variation of the metric in the Einstein frame \eqref{eq:AnalogueGR}. However, depending on the specific RBG theory chosen, it could become very difficult to express the gravity Lagrangian in terms of the matter fields.

\subsection{Mapping for Eddington-inspired Born-Infeld gravity }

For the sake of this paper we shall focus on EiBI gravity, a member of the RBG family whose action is given by \cite{BeltranJimenez:2017doy}
\begin{equation} \label{eq:ActEiBI}
\mathcal{S}_{EiBI}=\frac{1}{\epsilon \kappa^2} \int d^D x  \left(\sqrt{-\vert g_{\mu\nu}+\epsilon R_{\mu\nu} \vert}- \lambda\sqrt{-g} \right) \ ,
\end{equation}
where $\epsilon$ is a parameter with dimensions of length squared, and the theory features an effective cosmological constant given by $\Lambda_{eff}=\tfrac{\lambda-1}{\epsilon}$.
EiBI gravity is a particularly agreeable theory since the issue risen at the end of the previous section can be easily solved by noting that the EiBI Lagrangian density can be expressed as
\be
{\cal L}_{EiBI} = \frac{1}{\epsilon\kappa^2}\left[8\kappa^6\det\left(\frac{\delta{\cal L}_{EiBI}}{\delta R_{\mu\nu}}\right)-\lambda\right]\,, \label{eq:L-dLR}
\ee
eventually leading to the main problem of this procedure, namely, finding the functional dependence of $\frac{\delta{\cal L}_{EiBI}}{\delta R^{\mu}{}_{\nu}}$ with respect to the energy-momentum tensor. In the present case, using the definition of the auxiliary metric \eqref{eq:qDef}, expressed as
\be
q^{\mu\nu}=\frac{1}{4\kappa^4} \det{}^{-1}{\left(\frac{\delta{\cal L}_{EiBI}}{\delta R^{\rho}{}_{\sigma}}\right)}\frac{\delta{\cal L}_{EiBI}}{\delta R^{\lambda}{}_{\mu}}g^{\lambda\nu}\,,\label{eq:PreMetricMapGR-RBG}
\ee
it is easy to find the Ricci tensor in terms of the derivative of the gravity Lagrangian with respect to the Ricci tensor itself, namely
\be
\epsilon\,\tilde R^\mu{}_\nu = \delta^\mu_\nu - \left(2\kappa^2\right)^{-2} g^{-1}\det{}^{-1}{\left(\frac{\delta{\cal L}_{EiBI}}{\delta R_{\alpha\beta}}\right)}\frac{\delta{\cal L}_{EiBI}}{\delta R_{\mu\rho}}g_{\rho\nu}\ ,
\ee
so that the field equations \eqref{eq:AnalogueGR} can be recast as follows
\be
\det{}^{-1}{\left(\frac{\delta{\cal L}_{EiBI}}{\delta R^{\rho}{}_{\sigma}}\right)}\frac{\delta{\cal L}_{EiBI}}{\delta R^{\nu}{}_{\mu}} = 4\kappa^4\left[\delta^\mu_\nu - \epsilon\kappa^2\left(\tilde T^\mu{}_\nu-\tilde T \delta^\mu_\nu\right)\right]\ ,\label{eq:FieldEqsRecast}
\ee
leading to the general mapping between metrics by just replacing this last equation in \eqref{eq:PreMetricMapGR-RBG}:
\be
q^{\mu\nu}=\left[\delta^\mu_\lambda - \epsilon\kappa^2\left(\tilde T^\mu{}_\lambda-\tilde T \delta^\mu_\lambda\right)\right]g^{\lambda\nu}\,.\label{eq:MetricMapGR-RBG}
\ee
Furthermore, Eq.\eqref{eq:FieldEqsRecast} provides the determinant
\be
\det{\left(\frac{\delta{\cal L}_{EiBI}}{\delta R^{\rho}{}_{\sigma}}\right)} = \frac{1}{8\kappa^{6}}\det{}^{-1}{\left[\delta^\mu_\nu - \epsilon\kappa^2\left(\tilde T^\mu{}_\nu-\tilde T \delta^\mu_\nu\right)\right]}\,.
\ee
Using this last equation to simplify \eqref{eq:L-dLR}, one finally finds
\be
{\cal L}_{EiBI} = \frac{1-\lambda \sqrt{\det{\left[\delta^\rho_\sigma - \epsilon\kappa^2 \left(\tilde T^\rho{}_\sigma-\tilde T \delta^\rho_\sigma\right)\right]}}}{\epsilon\kappa^2\sqrt{\det{\left[\delta^\rho_\sigma - \epsilon\kappa^2 \left(\tilde T^\rho{}_\sigma-\tilde T \delta^\rho_\sigma\right)\right]}}}\,,\label{eq:GRMatterEiBI}
\ee
that can be substituted in \eqref{eq:RBGParamtr} to provide
\be
\mathcal{L}_m = \left.\frac{1}{\epsilon\kappa^2}\left\{\lambda-\frac{1-\epsilon\kappa^2\left(\tilde{\cal L}_m-\tilde T\right)}{\sqrt{\det{\left[\delta^\rho_\sigma - \epsilon\kappa^2 \left(\tilde T^\rho{}_\sigma-\tilde T \delta^\rho_\sigma\right)\right]}}}\right\}\right|_{q=q(g)}\,.\label{eq:RBG-Lag}
\ee

Let us point out that this is still a parameterization of the EiBI matter sector since the right-hand side is a function of the Einstein frame metric while we seek for a Lagrangian that fully depends on EiBI frame quantities. Instead of replacing the metric, we use the fact that, in the case of interest of this work, the matter content is specified to an electromagnetic field, implying that the matter Lagrangian depends on a single invariant that is written in terms of the field strength $F_{\mu\nu}=2\partial_{[\mu}A_{\nu]}$ as follows
\be
K=-\frac12 F_{\mu\nu}F^{\mu\nu}\,.
\ee
This is a consequence of the reduction property of arbitrary products of the field strength tensor and its dual according to the discussion in Appendix \ref{EMApp}. This observation leads to the intuition that the metric is always embedded in this invariant so that if we find a way to relate the invariants in different frames then the energy-momentum tensor in the Einstein frame could be written in terms of matter fields in the RBG frame. Following this strategy, we notice that the mapping equation \eqref{eq:MetricMapGR-RBG} can be inverted as follows
\begin{eqnarray}
g_{\mu\nu} &=& q_{\mu\nu} - \epsilon\kappa^2\left(\tilde T_{\mu\nu}- \tilde T q_{\mu\nu}\right) \label{eq:RBGMetr} \\
&=& \left[1+2\epsilon\kappa^2\left(\tilde{\cal L}_m - q^{\rho\sigma}\dfrac{\partial \tilde{\cal L}_m}{\partial q^{\rho\sigma}} \right)\right]q_{\mu\nu} + 2 \epsilon\kappa^2 \dfrac{\partial \tilde{\cal L}_m}{\partial q^{\mu\nu}}\ . \nonumber
\end{eqnarray}
Specifying this equation to the case of a free electromagnetic (Maxwell) field with a cosmological constant $\Lambda$ in the Einstein frame, namely $4\pi\tilde{\cal L}_m= \tilde K/2-4\pi\Lambda/\kappa^2$, with $\tilde K\equiv\frac12 q^{\mu\nu}F_{\mu\rho}q^{\rho\sigma}F_{\sigma\nu}$, we find
\begin{equation}
g_{\mu\nu} = \left[1-\epsilon\kappa^2 \left(\frac{\tilde K}{4\pi}+\frac{2\Lambda}{\kappa^2}\right)\right]q_{\mu\nu} +   \frac{\epsilon\kappa^2}{4\pi}\tilde K_{\mu\nu}  \ ,\label{Eq:EiBI-g-q}
\end{equation}
where we have introduced the tensor $\tilde K_{\mu\nu}\equiv \frac{\partial \tilde K}{\partial q^{\mu\nu}}$ and used the identity
\be
\frac{\delta\tilde{\cal L}_m}{\delta q^{\mu\nu}} = \frac{\partial\tilde{\cal L}_m}{\partial \tilde K}\tilde K_{\mu\nu}\,.
\ee
Using the following closure relation that follows from \eqref{eq:4F-Red}, namely
\be
K^\mu{}_\rho K^\rho{}_\nu = K \, K^\mu{}_\nu\,,\label{eq:KK}
\ee
we can easily calculate the inverse of \eqref{Eq:EiBI-g-q}, that is
\be
g^{\mu\nu} = \frac{1}{\left[1-\epsilon\kappa^2 \left(\frac{\tilde K}{4\pi}+\frac{2\Lambda}{\kappa^2}\right)\right]}\left(q^{\mu\nu} -  \frac{\epsilon\kappa^2}{4\pi(1-2\epsilon \Lambda)}\tilde K^{\mu\nu}\right)\,,
\ee
to finally find
\be
K=\frac12 g^{\mu\nu}F_{\mu\rho}g^{\rho\sigma}F_{\sigma\nu} = \tilde K\cdot \dfrac{(1-\tilde \epsilon \kappa^2 \tilde K)^2}{\left[1-\epsilon\kappa^2 \left(\frac{\tilde K}{4\pi}+\frac{2\Lambda}{\kappa^2}\right)\right]^2} \ ,
\ee
where we defined $\tilde \epsilon\equiv \epsilon/(4\pi(1-2\epsilon \Lambda))$. Note that the above equation reduces to the trivial relation between invariants  $
K = \tilde K\,,\label{eq:KtilKTilG}$ when the product $\epsilon \Lambda\to 0$, whereas in the general case (surprisingly) it simply leads to $K=\tilde K/(1-2\epsilon \Lambda)^2$.  With this result at hand, we can go back to the parametrization \eqref{eq:RBG-Lag} to see that the corresponding Lagrangian density in the RBG frame is given by
\be
\epsilon\kappa^2\mathcal{ L}_m=
	\lambda -
		\frac{1 - \epsilon\kappa^2\left(\frac{\tilde K}{4\pi}+\frac{2\Lambda}{\kappa^2}\right)
			}
		{\sqrt{\det\left\{
				\left[1 - \epsilon\kappa^2\left(\frac{\tilde K}{4\pi}+\frac{2\Lambda}{\kappa^2}\right)
				\right]\delta^\rho_\sigma +
				\frac{\epsilon\kappa^2}{4\pi} \tilde K^{\rho}{}_\sigma\right\}}}
	\nonumber \ .
\ee
Evaluating the determinant in the denominator of this expression using the formula
\be
\det{\left(A\delta^\mu_\nu + B K^\mu{}_\nu\right)} = A\left(A+B\,K\right)^2\,,
\ee
which follows from the composition rule \eqref{eq:KK}, and using that $K=\tilde K/(1-2\epsilon \Lambda)^2$  the RBG matter sector finally reads
\be
\mathcal{L}_m(K)=\displaystyle{
	\frac{1}{\epsilon\kappa^2}
	\left(\lambda - \sqrt{\frac{1}{1-2\epsilon \Lambda} - \frac{\epsilon\kappa^2}{4\pi} K}
	\right)}\ .\label{eq:Lm-final}
\ee
Note that the term $\epsilon \Lambda$ should be regarded as absolutely negligible because it involves the product of two tiny scales. Expanding in series that piece in the Lagrangian, we find
\be
\mathcal{L}_m(K)\approx
	\frac{1}{\epsilon\kappa^2}
	\left(\lambda - \sqrt{1 - 2\epsilon\kappa^2 \left(\frac{K}{8\pi}-\frac{\Lambda}{\kappa^2}\right)}
	\right) \ ,\label{eq:Lm-final}
\ee
which represents a Born-Infeld style version of the Einstein frame matter Lagrangian $\tilde{\mathcal{L}}_m=\frac{\tilde K}{8\pi}-\Lambda/\kappa^2$. This proves that the nonlinearities appearing in the EiBI gravity Lagrangian are somehow compensated by the nonlinear interactions of the electromagnetic field, giving rise to the 2+1 dimensional analogue of the Born-Infeld electrodynamics, where the Hodge dual field strength is absent since we are working in an odd dimensional theory.

\subsection{Alternative approach: anisotropic fluid representation}

An alternative and more pragmatic approach to the mapping of solutions between GR and RBG theories consists on using an anisotropic fluid description for electromagnetic fields, as first shown in \cite{Afonso:2018mxn}. We will consider this approach here too in order to double-check our results following a completely different route. To proceed, we first set an energy-momentum tensor in the RBG frame as
\begin{equation} \label{eq:Tmunug}
T_{\mu\nu}(g)=(\rho+p_{\perp})u_\mu u_\nu +p_{\perp} g_{\mu\nu} +(p_r-p_{\perp})\chi_\mu \chi_\nu  \ ,
\end{equation}
with time-like $u^{\mu}u^{\nu}g_{\mu\nu}=-1$ and space-like $\chi^{\mu}\chi^{\nu}g_{\mu\nu}=1$ vectors, and where $\rho$ is the energy density, $p_r$ the radial pressure and $p_{\perp}$ the tangential one. Similarly, we introduce another anisotropic fluid in the GR frame as
\begin{equation} \label{eq:Tmunuq}
\tilde{T}_{\mu\nu}(q)=(\tilde{\rho}+\tilde{p}_{\perp})v_\mu v_\nu +\tilde{p}_{\perp} q_{\mu\nu} +(\tilde{p}_r-\tilde{p}_{\perp})\xi_\mu \xi_\nu \ ,
\end{equation}
with new time-like $v^{\mu}v^{\nu}q_{\mu\nu}=-1$ and  space-like $\xi^{\mu}\xi^{\nu}q_{\mu\nu}=1$ vectors, and new density and pressures, ($\tilde{\rho},\tilde{p}_r,\tilde{p}_{\perp}$), characterizing the fluid. The fundamental relation between the matter sectors is given by \eqref{eq:PreCond}, that we rearrange as
\begin{equation} \label{eq:maptmunu}
\sqrt{-q}\,{\tilde{T}^\mu}\,_\n(q)= \sqrt{-g} \left[{T^\mu}_{\nu}(g)-\frac12 \delta^{\mu}_{\nu} \left(\mathcal{L}_G + T(g) \right) \right]\ ,
\end{equation}
and which can be recast in terms of the anisotropic fluid variables to obtain the correspondences
\begin{eqnarray}
\sqrt{-q}\,\tilde{p}_{\perp}&=&\sqrt{-g} \left(\frac{\rho +p_\perp - p_r - \mathcal{L}_G}{2} \right) \label{eq:map1} \\
\sqrt{-q}\left(\tilde{\rho}+\tilde{p}_{\perp}\right)&=& \sqrt{-g}\,\left(\rho+p_{\perp}\right) \label{eq:map2} \\
\sqrt{-q}\left(\tilde{p}_r-\tilde{p}_{\perp}\right)&=& \sqrt{-g}\,\left(p_r+p_{\perp}\right) \label{eq:map3}
\end{eqnarray}
relating the functions of the fluid on each frame. The relation between the determinant of the metrics is specified once a particular RBG theory is chosen, so that the mapping equations (\ref{eq:map1}), (\ref{eq:map2}) and (\ref{eq:map3}) allow to relate the matter sources on each frame. Particularizing to EiBI gravity, we can take advantage of the inverse of  (\ref{eq:MetricMapGR-RBG}) to see that
\begin{equation}
g_{\mu\nu}=(1+\epsilon \kappa^2 \tilde{T})q_{\mu\nu}-\epsilon \kappa^2 \tilde{T}_{\mu\nu} \ .
\end{equation}
As a result, given a metric $q_{\mu\nu}$ which satisfies Einstein's equations $G_{\mu\nu}(q)=\kappa^2 \tilde{T}_{\mu\nu}$, with $ \tilde{T}_{\mu\nu}$ as in (\ref{eq:Tmunuq}), then this $g_{\mu\nu}$ provides a solution to the EiBI equations with the $T_{\mu\nu}$ of (\ref{eq:Tmunug}) as source. In terms of the anisotropic fluid variables, we have that
\begin{eqnarray}
g_{\mu\nu}&=&[1+\epsilon\kappa^2(\tilde p_r-\tilde\rho)]q_{\mu\nu} - \epsilon\kappa^2(\tilde \rho + \tilde p_\perp) v_\mu v_\nu +  \nonumber \\
 &+&(\tilde p_r  - \tilde p_\perp)\xi_\mu \xi_\nu \ , \label{eq:gqgeneral}
\end{eqnarray}
which provides the $2+1-$dimensional solution of EiBI gravity once a seed solution in GR (as determined by $\{\tilde\rho,\tilde p_r,\tilde p_{\perp}\}$) is given. One of our tasks below will consist on identifying the fluid variables associated to the BTZ solution of GR in order to build a new solution in EiBI gravity.

\section{BTZ-type solutions in EiBI gravity}

\subsection{The BTZ solution of GR}

Following Carlip's notation \cite{Carlip:1995qv}, in a static, spherically symmetric 2+1 dimensional system of coordinates $(t,x,\phi$), the BTZ solution is given by
\begin{equation}
ds^2=-(N^{\perp})^2dt^2+f^{-2}dx^2+x^2(d\phi+N^{\phi}dt)^2 \ ,
\end{equation}
with the definitions
\begin{equation}
N^{\perp}=f=\left(-M+\frac{x^2}{l^2}+\frac{J^2}{4x^2} \right)^{1/2} ; N^{\phi}=-\frac{J^2}{2x^2} \ ,
\end{equation}
where $M$ is the ADM mass of the system, $l=-\Lambda^{-2}$ the AdS length and $J$ the angular momentum. To inspect the structure of this line element in more detail, it is instructive to explicitly rewrite it as
\begin{equation} \label{eq:lnbtz}
ds^2=\left(M-\frac{x^2}{l^2}\right)dt^2+f^{-2}dx^2 +x^2d\phi^2-Jdtd\phi \ ,
\end{equation}
which is (perhaps) a more canonical way of expressing an axially symmetric element. From it we can easily read off its physical structure. For instance, the single zero of the $g_{tt}$ component is found as
\begin{equation} \label{eq:erg}
x_{erg}=lM^{1/2} \ ,
\end{equation}
whose internal part defines the ergoregion, that is, the region inside which the observer is forced to co-rotate with the black hole, $d\phi/dt>0$. On the other hand, the Killing horizons of this geometry are given by the zeroes of $g^{xx}=f^2$, which are found as
\begin{equation} \label{eq:hor}
x_{\pm}^2=\frac{Ml^2}{2} \left(1\pm \left[1-\left(\frac{J}{Ml} \right)^2\right]\right) \ ,
\end{equation}
corresponding to the event and inner horizons, respectively. Using Eqs.(\ref{eq:erg}) and (\ref{eq:hor}) one can establish the different configurations in the BTZ solution. For positive masses, $M>0$, one finds a spectrum of black holes (covered by an ergohorizon) provided that $\vert J \vert \leq Ml$ (with the limiting case $J=Ml$ corresponding to extreme black holes), and naked singularities if $\vert J \vert > Ml$.  For null mass and angular momentum, $M=J=0$, a massless black hole, corresponding to purely AdS space, is found. And the holly crown of the BTZ solution corresponds to the $M=-1,J=0$ state, for which a regular de Sitter core is developed, disconnected from the black hole spectrum by a unit mass gap.

For the sake of this work we are interested in the generalization of the BTZ solution to the electrically charged case  coupled to a Maxwell field, $\tilde{\mathcal{L}}_m=\tilde K/2$. In such a case, the electromagnetic field is characterized by $A_\mu=(\frac{1}{2}Q \log\left(x/x_0\right),0,0)$, with $x_0$ an arbitrary length scale, and  Eq.(\ref{eq:lnbtz}) generalizes to
\begin{equation} \label{eq:BTZcanonical}
ds^2=-\left(H(x)-\frac{J^2}{4x^2}\right)dt^2+\frac{1}{H(x)}dx^2 +x^2d\phi^2-Jdtd\phi \ ,
\end{equation}
with  $H(x)=-M-\frac{Q^2}{2}\log (x/x_0)+\frac{x^2}{l^2}+\frac{J^2}{4x^2}$ and $Q$ the electric charge. The ergohorizons in this case, $g_{tt}=0$, are found as
\begin{equation} \label{eq:erghorbtz}
x_{\pm}= \pm\frac{lQ}{2}PL\left[\frac{-4e^{-4M/Q^2}}{l^2Q^2}\right]^{1/2} \ ,
\end{equation}
where $PL$ denotes the principal solution of Lambert's W function. This represents two real solutions depending on the combination of the values of $\{l, Q, M \}$, which means that there are branches of the solutions where the ergoregion is absent, similarly as in the usual four-dimensional  Kerr(-Newman) solution of GR. As for the horizons, $g^{xx}=H(x)$, no closed expression is possible but one can study the location of the extrema of $f^2$, resulting in
\begin{equation} \label{eq:horbtz}
x_{e}=l\left(Q^2 \pm \sqrt{Q^4+16J^2/l^2}/8\right)^{1/2} \ .
\end{equation}
In this way, writing $H(x_e)=H_e$ and assuming $H_e<0$  it can be easily shown that one has black holes with two horizons when $M>H_e$, extreme black holes if $M=H_e$, and naked singularities otherwise. The latter states also apply to all the $H_e>0$ cases. It is worth pointing out that in this electrically charged case the de Sitter $M=-1,J=0$ state of the uncharged case disappears, being replaced by $g^{xx}(x \approx 0) \approx -\tfrac{Q^2}{2}\log x$, thus corresponding to black holes with a single non-degenerate horizon. Therefore, the addition of charge to the BTZ solution destroys its most charismatic feature.

\subsection{Generating the EiBI-BTZ solution: electromagnetic approach}

Let us now head right to business. Mapping the charged BTZ solution into EiBI gravity is almost a trivial matter using equation (\ref{Eq:EiBI-g-q}) and knowing that the electromagnetic field in the Einstein frame is determined by $A_\mu=(\frac{1}{2}Q \log\left(x/x_0\right),0,0)$. With this $A_\mu$, one finds that
\begin{equation}
F_{\mu\nu}=\left(\begin{array}{ccc}  0 & -\frac{Q}{2x} & 0 \\ \frac{Q}{2x} & 0 & 0 \\ 0 & 0 & 0 \end{array} \right)  \ ,
\end{equation}
which leads to
\begin{equation}
\tilde{K}_{\mu\nu}=q^{\alpha\beta} F_{\mu\alpha}F_{\beta\nu}=\frac{Q^2}{4x^2}\left(\begin{array}{ccc} - f & 0 & 0 \\ 0 & \frac{1}{f} & 0 \\ 0 & 0 & 0 \end{array} \right)  \ ,
\end{equation}
and $\tilde{K}=q^{\mu\nu}\tilde{K}_{\mu\nu}/2=Q^2/4x^2$. As a result, we obtain
\begin{equation}
g_{\mu\nu}=\left[1-\epsilon\kappa^2\left(\frac{Q^2}{16\pi x^2}+\frac{2\Lambda}{\kappa^2}\right)\right]q_{\mu\nu}+\frac{\epsilon \kappa^2}{4\pi} \tilde{K}_{\mu\nu} \ ,
\end{equation}
and the line element turns into
\begin{eqnarray} \label{eq:BTZ_EiBI_EM}
\frac{ds^2}{\eta} &=&\left(M+\frac{Q^2}{2}\log \left(\frac{x}{x_0}\right)-\frac{x^2}{l^2} -\frac{\epsilon \kappa^2 J^2 Q^2}{64\pi \eta x^4}\right)dt^2 \nonumber \\
&+&H^{-1}dx^2 \\
&+&x^2\left(1-\frac{\epsilon \kappa^2 Q^2}{16\pi \eta x^2}\right)d\phi^2-J\left(1-\frac{\epsilon \kappa^2 Q^2}{16\pi \eta x^2}\right)dtd\phi \ , \nonumber
\end{eqnarray}
where we introduced $\eta\equiv (1-2\epsilon \Lambda)$. Note also that $\kappa^2=8\pi$ and $\Lambda=-1/l^2$.
This concludes our derivation of the line element in the EiBI gravity theory coupled to Born-Infeld electrodynamics.

\subsection{Generating the EiBI-BTZ solution: anisotropic fluid approach}

We will now derive again the line element obtained above but using the anisotropic fluid approach. In order to build the energy-momentum sourcing our three-dimensional rotating metric, which is the first step needed to generate the counterpart of the BTZ metric in EiBI gravity, we start with the line element (\ref{eq:BTZcanonical}) and introduce the normalized comoving (time-like) vector $v^\mu=(v^t,0,\omega v^t)$ (here $\omega$ denotes the angular velocity), which must satisfy
\begin{equation}
v^t=\pm \frac{2}{\sqrt{4J\omega +4H-\frac{J^2}{x^2}-4\omega^2 x^2}} \ .
\end{equation}
Similarly, the normalized radial (space-like) vector $\xi^{\mu}=(0,\xi^x,0)$ yields $\xi^x=H^{1/2}$. This way, the energy-momentum tensor of an anisotropic fluid in the GR frame, as given by Eq.(\ref{eq:Tmunuq}), is explicitly written as
\begin{equation}
{\tilde{T}^\mu}\,_\n(q)= \left(
\begin{array}{cccc}
\tilde{p}_{\perp} + \frac{(\tilde{\rho}+\tilde{p}_{\perp})(4Hx^2-J\tilde{J})}{\tilde{J}^2-4Hx^2} & 0  & \frac{2(\tilde{\rho}  + \tilde{p}_{\perp})x^2\tilde{J}}{\tilde{J}^2-4Hx^2} \\
0 & \tilde{p}_r & 0 \\
 \frac{(\tilde{\rho}+\tilde{p}_{\perp})\omega (4Hx^2-J\tilde{J})}{\tilde{J}^2-4Hx^2} & 0  & \tilde{p}_{\perp} + \frac{2(\tilde{\rho}+\tilde{p}_{\perp})\omega x^2\tilde{J}}{\tilde{J}^2-4Hx^2} \\
\end{array}
\right) \
\end{equation}
where $\tilde{J}=J-2\omega x^2$. Introducing the above expression into the right-hand side of the field equations (\ref{eq:AnalogueGR}) allows one to extract the different quantities characterizing  the fluid. In this sense, the component $(t\phi)$ of such field equations leads to $\omega=\tfrac{J}{2x^2}$, which inserted in the $(tt)$, $(xx)$, and $(\phi\phi)$ components yields the functions appearing in the energy-momentum tensor of the fluid as
\begin{equation}
\tilde{\rho}=-\tilde{p}_r=-\frac{J^2+2x^3H'}{4\kappa^2 x^4}; \tilde{p}_{\perp}=\frac{2x^4H''-3J^2}{4\kappa^2 x^4}
\end{equation}
Making explicit the $H(x)$ function of the BTZ solution, one finds
\begin{equation}
\tilde{\rho}=-\tilde{p}_r=\frac{l^2Q^2-4x^2}{4\kappa^2l^2 x^2} \hspace{0.1cm} ; \hspace{0.1cm} \tilde{p}_{\perp}=\frac{l^2Q^2+4x^2}{4\kappa^2l^2 x^2}
\end{equation}
which verifies that indeed we are dealing with a Maxwell field plus a cosmological constant as the matter sources of our setup. This corresponds to the BTZ solution described in the previous section. We have all thus ready to build the line element of EiBI gravity  coupled to the Born-Infeld-type electrodynamics (\ref{eq:Lm-final}) using the charged BTZ seed solution (\ref{eq:BTZcanonical}). Collecting all the definitions above, Eq.(\ref{eq:gqgeneral}) brings the result
\begin{eqnarray} \label{eq:mainresult}
\frac{ds^2}{\tilde{\eta}} &=&\left(M+\frac{Q^2}{2}\log \left(\frac{x}{x_0}\right)-\frac{x^2}{l^2} -\frac{\epsilon J^2 Q^2}{8\tilde{\eta} x^4}\right)dt^2 \nonumber \\
&+&H^{-1}dx^2  \\
&+&x^2\left(1-\frac{\epsilon Q^2}{2 \tilde{\eta}x^2}\right)d\phi^2-J\left(1-\frac{\epsilon Q^2}{2\tilde{\eta} x^2}\right)dtd\phi \ , \nonumber
\end{eqnarray}
where $\tilde{\eta}=(1+2\epsilon/l^2)$ coincides with our definition of $\eta$ upon the identification $\Lambda=-1/l^2$. Eq.(\ref{eq:mainresult}) exactly coincides with our previous result (\ref{eq:BTZ_EiBI_EM}) and represent collectively the main result of this work. As we have proved in previous sections, this solution corresponds to an axially symmetric, electrically charged object with a cosmological constant in the EiBI gravity theory coupled to a Born-Infeld-type electrodynamics system in $2+1$ space-time dimensions. In the next section we shall proceed with the analysis of its most physically salient features.

Before going into that, let us consider that given the fact that the conformal factor $\tilde{\eta}$ is essentially unity and does not play any relevant role in the physics because it can be reabsorbed into a global redefinition of units, from now on we will omit it in our discussion. We will also use the notation $\tilde{\epsilon}\equiv \epsilon/\tilde{\eta}$ to lighten the notation. In addition, since $\epsilon$ represents a small length squared and $l^2$ is a large length scale squared, we will assume throughout that $\tilde \eta>0 $ everywhere regardless of the sign of $\epsilon$.

\section{Structure of the EiBI-BTZ solution\label{SolutionStructure}}

\subsection{Radial function and wormhole structures}

The first noticeable difference of the EiBI solution (\ref{eq:mainresult}) with respect to the charged BTZ one  (\ref{eq:BTZcanonical}) lies on the behaviour of the radial coordinate in the component $g_{\phi\phi}$. Indeed, in our EiBI solution the latter reads as
\begin{equation} \label{eq:rsx}
r^2(x)=x^2-s x_c^2\ ,
\end{equation}
where we have defined
\begin{equation} \label{eq:minsp}
x_c^2=\tfrac{ \vert \tilde \epsilon \vert Q^2}{2} \ .
\end{equation}
Note that here $r^2(x)>0$ is the areal (circumference) radius and that the domain of definition of  the coordinate $x$ depends on the sign $s=\pm1$ of $\tilde\epsilon$, which  has a non-trivial impact on the structure of the corresponding solutions. In order to analyze this aspect, we introduce the notation $\tilde{\epsilon}=s \vert \tilde \epsilon \vert$ and split the discussion into the subcases $s=+1$ and $s=-1$.

\begin{figure}[t!]
	\centering
	\includegraphics[width=4.0cm,height=4.0cm]{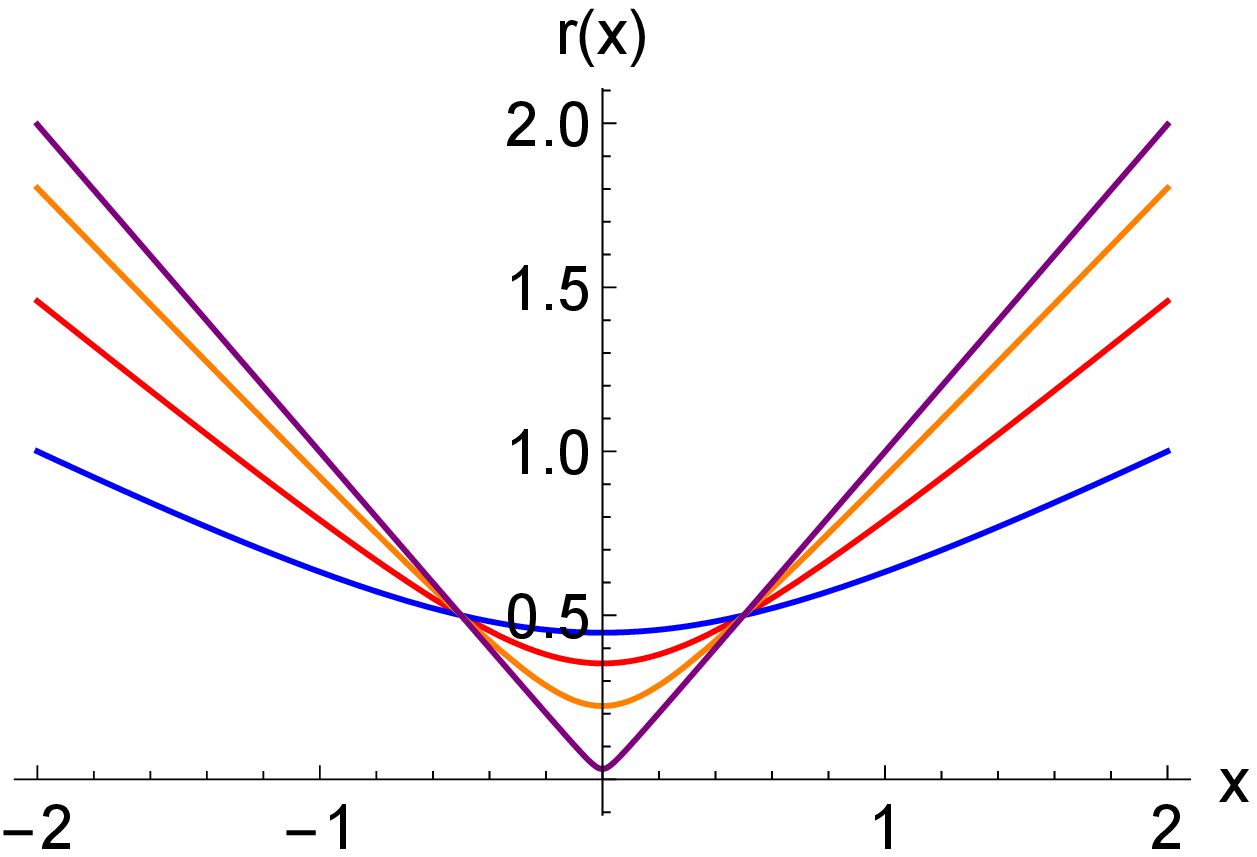}
	\includegraphics[width=4.0cm,height=4.0cm]{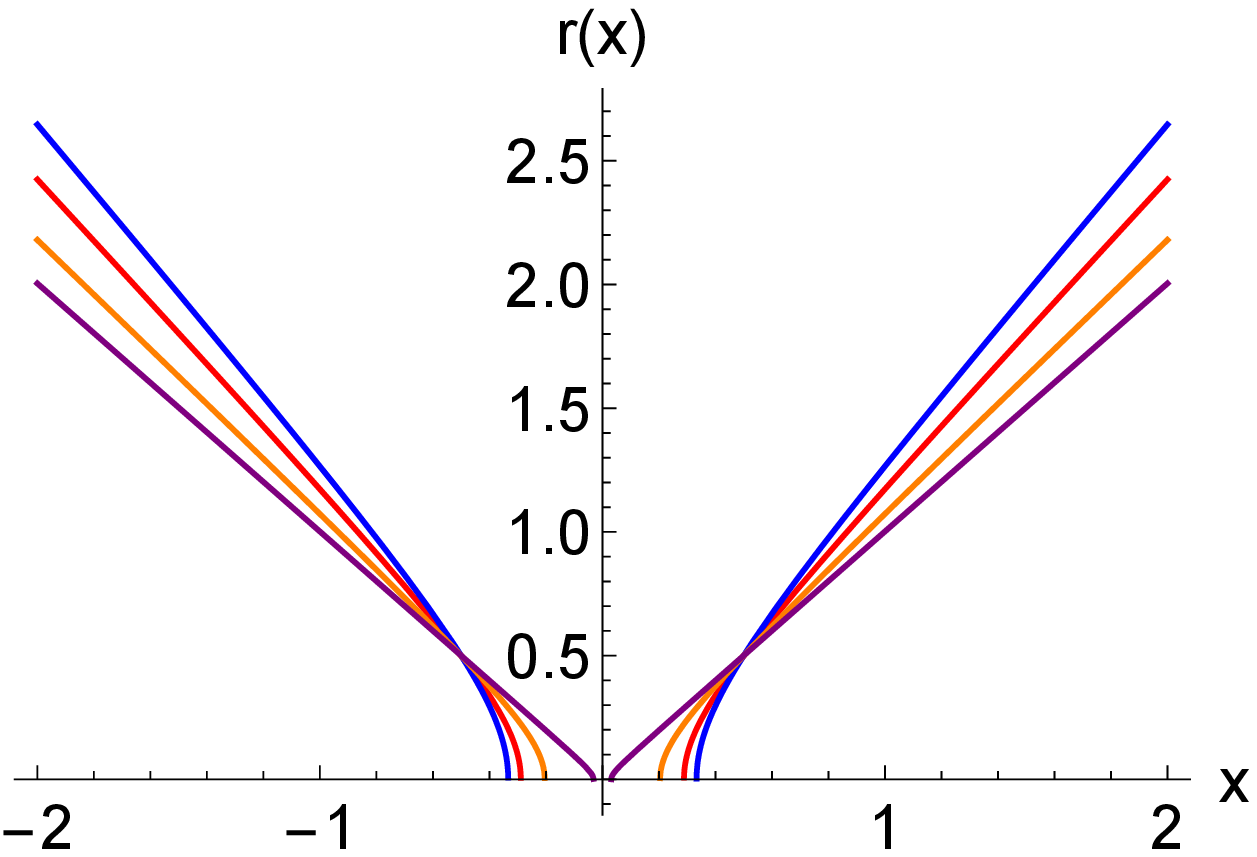}
	\caption{Behaviour of the radial function $r(x)$ in Eq.(\ref{eq:rsx}) for the cases $s=-1$ (left) and $s=+1$ (right). In this plot we have taken $Q=l^2=1$ and four cases of $\vert \epsilon \vert=1/500$ (purple), $\vert \epsilon \vert=1/10$ (orange), $\vert \epsilon \vert=1/4$ (red), and $\vert \epsilon \vert=0.4$ (blue).  }
	\label{fig:wh}
\end{figure}

In the $s=-1$ case one finds that the radial function $r^2(x)$ in Eq.(\ref{eq:rsx}) has a minimum circumference of radius $L_{min}=2\pi x_c$ at $x=0$. At this point the radial function $r^2(x)$ bounces off and the solution is naturally extended to the full range  $x\in ]-\infty,+\infty[$ (due care should be taken with the $\log(x/x_0)$ function in the metric, which should be regarded as $\log|x/x_0|$ if the negative $x$ domain is considered). This is the typical behaviour of a wormhole structure, as depicted in Fig. \ref{fig:wh}, with $x=0$ ($r=x_c$) representing its throat. The latter is circularly symmetric despite the axial symmetry of the space-time, since the contributions of the angular momentum only appear in the $g_{tt}$ and $g_{t\phi}$ components of the line element. When $\tilde\epsilon \to 0$ one has that $r^2(x) \approx x^2$ and the wormhole mouth closes, recovering the point-like singularity of the charged BTZ solution.

As for the $s=+1$ branch, one finds that a zero in the radial function takes place at $x=x_c$,
so that the space-time would split into two disconnected regions $x\in(-\infty,-x_c) \bigcup x \in (x_c, +\infty) $ which are causally disconnected from one another. In fact, the region within $|x|<x_c$ is forbidden because it would imply a change of signature in the metric. Also at this point the  $g_{t\varphi}$ component vanishes, indicating that the effective angular velocity vanishes as this center is approached.

\subsection{Structure of the line element: ergospheres and horizons}

\begin{figure}[t!]
	\centering
	\includegraphics[width=8.0cm,height=5.5cm]{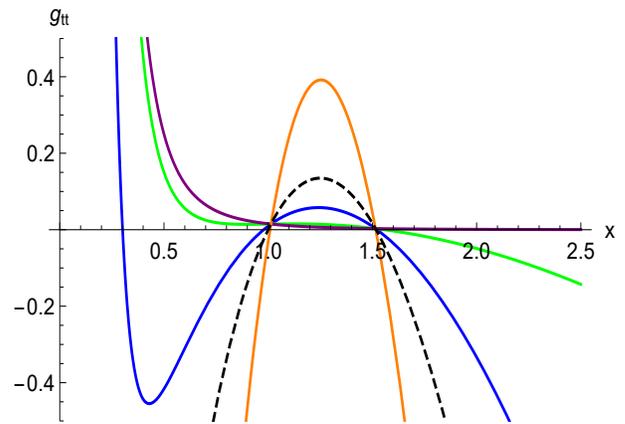}
	\caption{The behaviour of the metric component $g_{tt}$ for the parameters $\{l=1$, $M=1$, $Q=2.5$, $J=0.2\}$ in the cases $s=-1$ with $\epsilon=-0.3$ (blue), $\epsilon=-0.47$ (green), $\epsilon=-0.5$ (purple), $\epsilon=+1$ (orange) and its comparison with the GR case $\epsilon=0$ (dashed black). Despite these numbers being quite unreasonable, since the charge should be not too large as compared to the mass, while the length $l^2$ should be very large as compared to $M$ to avoid having a very small universe, these numbers are chosen to highlight the existence of additional configurations in terms of ergospheres as compared to the usual BTZ solutions of GR.}
	\label{fig:gtt}
\end{figure}

Let us now study the structure of horizons of this solution. We first note that the asymptotic structure of the component $g_{tt}$ is given by
\begin{equation}
g_{tt}(x \approx \infty) \sim -\frac{x^2}{l^2}  +\left( M +\frac{ Q^2}{2} \log(x/x_0)\right) + \mathcal{O}(1/x^2)
\end{equation}
being asymptotically Anti-de Sitter. Regarding the existence of an ergoregion, one needs to look for the zeroes of $g_{tt}$, which are given by the solutions of the trascendental equation
\begin{equation}\label{eq:ergoregions}
\frac{x^2}{l^2}-\frac{Q^2}{2}\left(\log(x/x_0)-\frac{\tilde{\epsilon} J^2}{4x^4}\right)-M=0 \ ,
\end{equation}
where, as opposed to the ergohorizons of the original charged BTZ case, Eq.(\ref{eq:erghorbtz}), now the contribution of a crossed term involving the angular momentum and the electric charge pops us modulated by the EiBI parameter $\tilde{\epsilon}$. Therefore, in the limit $J=0$ one gets exactly the same ergohorizons as in the GR case, while for $J \neq 0$ no exact solutions of this equation can be found.  For $J\neq 0$ the analysis must be split into a case with $M>0$ and another with $M<0$. Assuming that $l^2$ is the largest length scale involved, if $M>0$ then $g_{tt}$ has a zero in the large $x$ limit when $x^2\approx M l^2$, while for $M<0$ there is no such zero.  In the opposite limit, $x\to 0$, the sign of $\tilde\epsilon$ becomes relevant. If $|\tilde\epsilon|>0$ then Eq. (\ref{eq:ergoregions}) has no zeros as $x\to 0$. However, if $\tilde\epsilon<0$ then there is a competition between the two divergent terms multiplied by $Q^2$ because $\lim_{x\to 0}\log(x/x_0)\to -\infty$ but $\lim_{x\to 0} -\frac{\tilde{\epsilon} J^2}{4x^4} \to +\infty$, which introduce a minimum in $g_{tt}$. Since in this limit the $x^2/l^2$ term is negligible and $M$ is just a constant, the key element to see if there are zeros in $g_{tt}$ is the location of the minimum defined by the $Q^2$ terms. This minimum is located at $x^4_m=J^2|\tilde\epsilon|$, which implies $g_{tt}\vert_{x_m}\approx M +Q^2\left(1+\log\left(\frac{J^2|\tilde\epsilon|}{x_0^4}\right)\right)/8$. If this quantity is negative, then $g_{tt}$ will have two additional zeros in the neighborhood of $x_m$ while it will have no new zeros if it is positive. We thus conclude that if $\tilde\epsilon<0$ and $M>0$, we can have up to three ergohorizons [see Fig. \ref{fig:gtt}]. It is not difficult to see that if $M<0$ then the large $x$ zero disappears and the maximum number of ergohorizons will be limited to just two. If we focus now on the case $\tilde\epsilon>0$, then $M>0$ allows for a large $x$ zero due to the $l^2$ term, and there is another zero for small values of $x$ due to the fact that the $Q^2$ terms contribute with the same sign but opposite to $M$. If $M<0$, then only one zero would be possible.

To find the horizons of these configurations, one computes the norm of the vectors normal to the $x=$ constant hypersurface, which yields
\begin{equation}
\xi_{\mu} \xi^{\mu}=g^{xx}=H(x) \ ,
\end{equation}
and, therefore, the Killing horizons for this geometry are given by $H(x)=0$, which is the same result as in GR. Therefore, the discussion of the structure of horizons is exactly the same as in GR.

\subsection{Geodesic completeness and regularity}

Let us now deal with the issue of the regularity of these solutions. We recall that in the charged BTZ solution of GR the curvature scalar and the Kretschmann are in general divergent at the center, $x=0$. In particular, for the line element (\ref{eq:BTZcanonical}) they are given by
\begin{equation}
R_{\text{GR}}  = \frac{-6}{l^2} + \frac{Q^2}{2x^2}\hspace{0.1cm};\hspace{0.1cm}K_{\text{GR}} =\frac{12}{l^4} +\frac{3Q^4}{4x^4}-\frac{2Q^2}{l^2x^2} \ ,
\end{equation}
where we see that the divergent pieces are due to the electric field contribution. When the charge is absent, one finds the well known regularity of the vacuum BTZ solutions. Let us now carry out the same exercise for the EiBI-BTZ solutions (\ref{eq:mainresult}). Focusing on the  $s=-1$ configurations, one finds the expansion of the curvatures at the wormhole throat, $x=0$, as
\begin{eqnarray}
R_{\text{EiBI}} &\approx & \frac{\vert \tilde\epsilon \vert J^2Q^2}{4 x^6} + \mathcal{O}(1/x^4)  \\
K_{\text{EiBI}} &\approx & \frac{11 \vert \tilde\epsilon \vert^2 J^4 Q^4}{16 x^{12}} + \mathcal{O}(1/x^{10}) \ ,
\end{eqnarray}
which can be softened down to $ \mathcal{O}(1/x^2)$ and $\mathcal{O}(1/x^6)$ when $J=0$, but can never go away, similarly as in the charged BTZ solution of GR.
In the $s=+1$ case, we find instead
\begin{eqnarray}
R_{\text{EiBI}} &\approx & \frac{H(x_c)}{2 (x-x_c)^2} + \mathcal{O}(1/ (x-x_c))  \\
K_{\text{EiBI}} &\approx & \frac{H(x_c)^2}{4 (x-x_c)^4} + \mathcal{O}(1/ (x-x_c)^3) \ ,
\end{eqnarray}
where $x_c$ was defined in (\ref{eq:minsp}). We thus see that regardless of the sign of $\epsilon$, these solutions always have divergences represented by poles of, at least, the same order as in GR.

To rigorously discuss the regularity (or not) of the above configurations, we need to call upon the analysis of geodesic completeness \cite{Senovilla:2014gza}. This is so because, being time-like and null geodesics associated to the free-falling motion of physical observers and to the transmission of information, respectively, the requirement of their completeness is a basic requirement for the predictability of the corresponding theory.
This analysis can be carried out by attending to the existence of two conserved quantities, $E=-g_{\mu\nu}\xi^\mu u^\nu$ and $L=g_{\mu\nu}\Phi^\mu u^\nu$, where $\xi^\mu=(1,0,0)$ represents the time Killing vector $\partial_t$ and $\Phi^\mu=(0,0,1)$ is the angular Killing vector $\partial_\phi$, and also to the fact that $c=-g_{\mu\nu}u^\mu u^\nu$ is $c=0,1$ for null and time-like geodesics with tangent vector $u^\mu$. With these three constants of motion one can derive three equations that determine the evolution of the tangent vector $u^\mu$ \cite{Carter:1968ks}. From the conserved charges, we find
\begin{eqnarray}
\frac{dt}{Ed\lambda}&=& \frac{1}{H(x)} \left(1-\frac{b J}{2 x^2}\right) \ , \\
\frac{d\phi}{E d\lambda}&=& \frac{1}{H(x)}
\left(
\frac{J}{4 x^2}+
b\frac{ \left(H(x)-\frac{J^2}{4x^2}\left[1-s \frac{x_c^2}{x^2}\right]\right)}
{x^2-s x_c^2}
\right) \ ,
\end{eqnarray}
and combining them with $0=g_{\mu\nu}u^\mu u^\nu$ we get
\begin{eqnarray}
\left(\frac{dx}{Ed\lambda}\right)^2&=& 1-\frac{b J}{x^2}- \frac{b^2\left(H(x)-\frac{J^2}{4x^2}\left[1-s \frac{x_c^2}{x^2}\right]\right)}{x^2-s {x_c^2}} \ .
\end{eqnarray}
In the above formulas we have defined the impact parameter $b\equiv L/E$. We will not provide a detailed analysis of the curves that follow from these equations,  but rather we shall focus on their properties near $x=0$ when $\tilde\epsilon<0$ (wormhole case) and near $x_c$ when $\tilde\epsilon>0$, where the potentially dangerous regions lie.

Considering first the case  $\tilde\epsilon\equiv -2l_\epsilon^2<0$, in the limit $x\to 0$ the relevant  equations with $J\neq 0$ boil down to
\begin{eqnarray}
\frac{d\phi}{E d\lambda}&\approx &  -\frac{b}{x^2} \ , \\
\left(\frac{dx}{Ed\lambda}\right)^2&\approx & \frac{b^2J^2}{4 x^4} \ ,
\end{eqnarray}
whereas for the non-rotating case $J=0$ we have instead
\begin{eqnarray}
\frac{d\phi}{E d\lambda}&\approx &  \frac{b}{l_\epsilon^2Q^2} \ , \\
\left(\frac{dx}{Ed\lambda}\right)^2&\approx & 1-b^2\frac{\left(M+\frac{Q^2}{2}\log\left(\frac{x}{x_0}\right)\right)}{l_\epsilon^2Q^2} \ . \label{eq:dxdl-J0}
\end{eqnarray}
It is easy to see that for $J\neq 0$ one can reach $x=0$ in a finite amount of proper time. Note also that though the angular velocity $d\phi/d\lambda$ and the radial velocity $dx/d\lambda$ diverge at that point, the ratio $d\phi/dx=-2/J$ is well defined as long as $J\neq 0$.  On the other hand, when $J=0$ the sign of $M$ becomes relevant. If $M>0$, then some light trajectories could bounce at some $x>0$ (depending on model parameters), while if $M<0$ then in the $x\to 0$ limit the right-hand side of (\ref{eq:dxdl-J0}) will always be positive definite and dominated by
\begin{equation}
\left(\frac{dx}{Ed\lambda}\right)^2 \approx  -\frac{b^2}{2l_\epsilon^2}\log\left(\frac{x}{x_0}\right) \ ,\end{equation}
which also implies that $x\to 0$ is reached in a finite amount of proper time (this can be checked explicitly via numerical integration). Together with the existence of divergences in the curvature scalars and the line element as $x\to 0$, the fact of reaching  $x\to 0$ in finite affine time strongly suggests that the space-time is singular at that point regardless of the value of $J$.

Consider now the case  $\tilde\epsilon>0$. In the limit $x\to x_c$, the relevant  equations boil down to
\begin{eqnarray}
\frac{d\phi}{E d\lambda}&\approx &  \frac{b}{2 x_c (x- x_c)} \ , \\
\left(\frac{dx}{Ed\lambda}\right)^2&\approx &\frac{b^2 H(x_c)}{2 x_c (x- x_c)} \ , \label{eq:limEpos}
\end{eqnarray}
Note that here, unlike in the $\tilde\epsilon<0$ case, the angular momentum does not play any crucial role beyond its influence on the sign of the right-hand side of (\ref{eq:limEpos}). Focusing on those geodesics that can reach the center,  namely, those for which $H(x_c)>0$, the above limits show that $x\to x_c $ is always reached in finite affine time, with $\phi_c\approx \phi_0+\sqrt{2(x-x_c)/x_c H(x_cn)}$ approaching the constant finite value $\phi_0$. For the same reasons as for $\tilde\epsilon<0$, this case is as singular as in GR.

\section{Conclusion}

The main aim of this work was the construction of an exact solution for a charged BTZ-type configuration with  cosmological constant term within the context of an extension of General Relativity dubbed as Eddington-inspired Born-Infeld gravity. This has been achieved thanks to the recent development of a powerful method to map known solutions of General Relativity with some matter source into other metric-affine gravity theories coupled to a modified matter source. Via this method it is possible to find a solution of the latter out of a seed solution of the former via purely algebraic transformations, rather than directly solving the complicated non-linear structure of the modified gravity field equations.

Here we have explicitly shown that the application of this method to GR coupled to an electromagnetic (Maxwell) field in $2+1$ dimensions is mapped into EiBI gravity coupled to a Born-Infeld-type electrodynamics. We have been able to obtain the line element of the modified theory following two different approaches, namely, i) focusing entirely on the electromagnetic fields, and ii) considering an effective anisotropic fluid description of the matter sources. The obtained results are coincident and provide a nontrivial consistency check of the mapping relations and of their implementation.

The discussion of the properties of the obtained solutions has been subsequently split into two cases, depending on the sign $s$ of the EiBI parameter, $s=\epsilon/ \vert \epsilon \vert =\pm 1$.
We found that in the $s=-1$ case the (areal) radial function has a minimum at $x=0$ ($r=x_c>0$), representing the typical behavior of a wormhole structure. The wormhole may be covered by two horizons or be naked, just like in the original BTZ solutions of GR, while more than one ergohorizon may be present depending on model parameters. Despite the presence of a wormhole, which in some cases can cure certain pathologies, we found curvature divergences at the throat with a pole structure stronger or equal to that found in the GR solution. Together with the fact that nothing prevents null geodesics from reaching this region in finite affine time, everything suggests that the resulting space-time remains as singular as in the charged BTZ solution of GR. Something similar occurs in the $s=+1$ case, though here curvature invariants diverge at a finite value of the radial coordinate, $x=x_c$, where the radial function vanishes, thus representing the center of the solution.

The results presented in this paper are yet another proof of the usefulness of the mapping method in finding exact analytical solutions of theoretical and/or observational interest in modified gravity theories.
We hope to further report on the applications of this method in physically more appealing scenarios in $2+1$ and higher dimensions, like, for instance, the direct coupling of EiBI  gravity or other gravity theories to Maxwell electrodynamics, which should be in correspondence with solutions to Einstein's equations coupled to nonlinear theories of electrodynamics. Work in this direction is currently underway.



\section*{Acknowledgments}

MG is funded by the predoctoral contract 2018-T1/TIC-10431. DRG is funded by the \emph{Atracci\'on de Talento Investigador} programme of the Comunidad de Madrid (Spain) No. 2018-T1/TIC-10431, and acknowledges further support from the Ministerio de Ciencia, Innovaci\'on y Universidades (Spain) project No. PID2019-108485GB-I00/AEI/10.13039/501100011033, and the FCT projects No. PTDC/FIS-PAR/31938/2017 and PTDC/FIS-OUT/29048/2017. This work is supported by the Spanish projects  FIS2017-84440-C2-1-P and PID2020-116567GB-C21 (MINECO/FEDER, EU), the project PROMETEO/2020/079 (Generalitat Valenciana), and the Edital 006/2018 PRONEX (FAPESQ-PB/CNPQ, Brazil, Grant 0015/2019). This article is based upon work from COST Action CA18108, supported by COST (European Cooperation in Science and Technology).

\appendix

\section{Basic properties of the electromagnetic field\label{EMApp}}

The problem that we address in this appendix is how to find electromagnetic invariants of the electromagnetic field in a 3 dimensional space-time. In such a case, the electric field is specified by a two components vector while the magnetic field has just one component. This is consistent with the Hodge dual of the field strength tensor $F_{\mu\nu}$, which is just a vector, ${}^\star F^\mu = \tfrac{\sqrt{-g}}{2} \epsilon^{\mu\nu\rho}F_{\rho\sigma}$. These are the tensorial quantities that we can combine to build Lorenz invariant quantities. Underestanding how arbitrary products of $F_{\mu\nu}$ and ${}^\star F_\mu$ can be simplified is fundamental to find a minimal set of invariants. To this end, notice that the product of four field strengths can be reduced according to
\begin{equation}
F^\mu{}_\nu F^\nu{}_\rho F^\rho{}_\sigma F^\sigma{}_\gamma = -\frac12 F_{\rho\sigma}F^{\rho\sigma} F^\mu{}_\nu F^\nu{}_\gamma - F^\mu{}_\nu F^\nu{}_\rho {}^\star F^\rho {}^\star F_\gamma\,,\label{4F}
\end{equation}
where the following identity has been used
\begin{eqnarray}
F^\rho{}_\sigma F^\sigma{}_\gamma &=& {}^\star F_{\alpha}{}^\star F^{\alpha} \delta^\rho{}_\gamma - {}^\star F^\rho {}^\star F_\gamma \nonumber \\
&=& -\frac12 F_{\alpha\beta}F^{\alpha\beta} \delta^\rho{}_\gamma - {}^\star F^\rho {}^\star F_\gamma \ , \label{2F}
\end{eqnarray}
Focusing on the last term of \eqref{4F}, one finds that
\begin{equation}
F^\mu{}_\nu F^\nu{}_\rho {}^\star F^\rho = \left({}^\star F_{\alpha}{}^\star F^{\alpha} \delta^\mu{}_\rho - {}^\star F^\mu {}^\star F_\rho\right) {}^\star F^\rho = 0 \ ,
\end{equation}
that reduces Eq.\eqref{4F} to
\begin{equation}
F^\mu{}_\nu F^\nu{}_\rho F^\rho{}_\sigma F^\sigma{}_\gamma = -\frac12 F_{\rho\sigma}F^{\rho\sigma} F^\mu{}_\nu F^\nu{}_\gamma \ . \label{eq:4F-Red}
\end{equation}
This relation has importance consequences since it states that arbitrary products of field strengths can be written in terms of powers of the simplest invariant $F_{\rho\sigma}F^{\rho\sigma}$ multiplied by the product of just two field strengths. Contrary to the four dimensional case, this reduction process does not involve the magnetic field. In particular, the following relation holds
\begin{equation}
F^{\lambda_1}{}_{\lambda_2} F^{\lambda_2}{}_{\lambda_3}\cdots F^{\lambda_{2n}}{}_{\lambda_1} =  \left({}^\star F_{\alpha}{}^\star F^{\alpha}\right)^n = \frac{\left(-1\right)^n}{2^n} \left(F_{\mu\nu}F^{\mu\nu}\right)^n \,. \label{4F-RedSat}
\end{equation}
Throwing into the game ${}^\star F^\mu$ does not lead to a radical difference. In fact, the relevant tensorial quantities to consider are reduced as follows
\begin{eqnarray}
&&F^{\mu}{}_{\lambda_1} F^{\lambda_1}{}_{\lambda_2}\cdots F^{\lambda_{2n-1}}{}_{\lambda_{2n}} {}^\star F_{\mu} {}^\star F^{\lambda_{2n}} \label{4F-RedSat} \\
&=& -\frac{\left(-1\right)^n}{2^{n-1}} \left(F_{\mu\nu}F^{\mu\nu}\right)^{n-1} F^{\mu}{}_{\lambda_1}F^{\lambda_{1}}{}_{\lambda_{2n}} {}^\star F_{\mu} {}^\star F^{\lambda_{2n}}=0  \nonumber  \,,
\end{eqnarray}
and
\begin{eqnarray}
&&F^{\mu}{}_{\lambda_1} F^{\lambda_1}{}_{\lambda_2}\cdots F^{\lambda_{2n}}{}_{\lambda_{2n+1}} {}^\star F_{\mu} {}^\star F^{\lambda_{2n+1}}  \label{4F-RedSat}\\
&=& -\frac{\left(-1\right)^n}{2^{n-1}} \left(F_{\mu\nu}F^{\mu\nu}\right)^{n-1} F^{\mu}{}_{\lambda_1}F^{\lambda_{1}}{}_{\lambda_{2n}}F^{\lambda_{2n}}{}_{\lambda_{2n+1}} {}^\star F_{\mu} {}^\star F^{\lambda_{2n+1}} \nonumber \\
&=&0 \nonumber  \ ,
\end{eqnarray}
where the last implication is a consequence of \eqref{2F} since
\begin{eqnarray}
&&F^{\mu}{}_{\lambda_1}F^{\lambda_{1}}{}_{\lambda_{2n}}F^{\lambda_{2n}}{}_{\lambda_{2n+1}} {}^\star F_{\mu} {}^\star F^{\lambda_{2n+1}}  \label{4F-RedSat} \\
&=& \left({}^\star F_{\alpha}{}^\star F^{\alpha} \delta^\mu_{\lambda_{2n}} - {}^\star F^{\mu}{}^\star F_{\lambda_{2n}}\right)F^{\lambda_{2n}}{}_{\lambda_{2n+1}} {}^\star F_{\mu} {}^\star F^{\lambda_{2n+1}}=0 \,. \nonumber
\end{eqnarray}
We therefore deduce that there is only one independent invariant build with the field strength and its dual. The situation would change when the vector potential is included in the construction since the the number of invariants grows considerably.

\end{document}